\pdfoutput=1
\documentclass[titlepage, 12pt]{article}

\usepackage{asa}
\usepackage[]{natbib}
\bibpunct{(}{)}{;}{a}{}{,}

\usepackage{amsmath, amsthm, amsbsy, epsfig, epsf, psfrag, alltt, booktabs, graphicx, amssymb, placeins, rotating, subfig, mathrsfs}

\usepackage[margin=1in]{geometry}

\usepackage{authblk}

\newcommand\ceil[1]{\lceil#1\rceil}

\graphicspath{{./figs/}}
\DeclareGraphicsExtensions{.eps,.ps,.pdf}
\begin{document}

\title{Automated Threshold Selection for Extreme Value Analysis via
    Goodness-of-Fit Tests with Application to Batched Return Level Mapping
  \thanks{Research partially supported by NSF grant
  DMS 1521730, University of Connecticut Research Excellence
  Program, and Environment Canada.
  The authors thank Prof. Vartan Choulakian for the discussion and
  insight on approximating the tails of the null distribution of the
  Anderson--Darling  and Cram\'{e}r--von Mises statistics for generalized 
  Pareto distributed data.}}

\renewcommand\Affilfont{\normalfont\small}
\author[1]{Brian Bader}
\author[1]{Jun Yan}
\author[2]{Xuebin Zhang}
\affil[1]{Department of Statistics, University of Connecticut}
\affil[2]{Environment and Climate Change Canada}

\maketitle

\begin{abstract}
Threshold selection is a critical issue for extreme value analysis 
with threshold-based approaches. Under suitable conditions, 
exceedances over a high threshold have been shown to
follow the generalized Pareto distribution (GPD) asymptotically.
In practice, however, the threshold must be chosen. If the chosen 
threshold is too low, the GPD approximation may not hold and 
bias can occur. If the threshold is chosen too high, reduced 
sample size increases the variance of parameter estimates. To 
process batch analyses, commonly used selection methods such 
as graphical diagnosis are subjective and cannot be automated, 
while computational methods may not be feasible.
We propose to test a set of thresholds through the goodness-of-fit
of the GPD for the exceedances, and select the lowest 
one, above which the data provides adequate fit to the GPD. 
Previous attempts in this setting are not valid due to the special 
feature that the multiple tests are done in an ordered fashion.
We apply two recently available stopping rules that control the 
false discovery rate or familywise error rate 
to ordered goodness-of-fit tests to automate threshold selection.
Various model specification tests such as the Cram\'{e}r--von Mises, 
Anderson--Darling, Moran's, and a score test are investigated.
The performance of the method is assessed in a large scale  
simulation study that mimics practical return level estimation.
This procedure was repeated at hundreds of sites in the western US 
to generate return level maps of extreme precipitation.

\vspace{1cm}

\end{abstract}

\section{Introduction}
\label{s:intr}

Extreme value analysis has wide applications in a variety of fields,
such as hydrology \citep[e.g.,][]{katz2002statistics} 
and climatology \citep[e.g.,][]{kharin2007changes, 
kharin2013changes, wang2008downscaling}. 
Return levels, the levels of a measure of interest that is expected 
to be exceeded on average once every certain period of time
(return period), are a major goal of inference in these fields.
Commonly used in extreme value analysis, threshold-based 
methods involve modeling data exceeding a suitably chosen 
high threshold with the generalized Pareto distribution (GPD)
\citep{balkema1974residual, pickands1975statistical}.
Choice of the threshold is critical in obtaining accurate 
estimates of model parameters and return levels.
The threshold should be chosen high enough for the 
excesses to be well approximated by the GPD to minimize bias, 
but not so high to substantially increase the variance of the estimator 
due to reduction in the sample size (the number of exceedances).

Although it is widely accepted in the statistics community that 
the threshold-based approach, in particular peaks-over-threshold (POT), 
use data more efficiently than the block maxima method 
\citep[e.g.,][]{caires2009comparative}, it is much less used than the
block maxima method in some fields such as climatology. 
The main issue is the need to conduct the analyses over many
locations, sometimes over hundreds of thousands of 
locations~\citep[e.g.,][]{kharin2007changes, kharin2013changes}, and
there is a lack of efficient procedures that can automatically select
the threshold in each analysis. 
For example, to make a return level map of annual maximum daily 
precipitation for three west coastal US states of California, Oregon, 
and Washington alone, one needs to repeat the estimation procedure
including threshold selection,  at each of the hundreds of stations.
For the whole US, thousands of sites need to be processed.
A graphical based diagnosis is clearly impractical. 
It is desirable to have an intuitive automated threshold selection
procedure in order to use POT in analysis.

% discuss previous work on selection methods (single site analysis for now)
Many threshold selection methods are available in the literature;
see \citet{scarrott2012review} and \citet{caeiro2016threshold} 
for recent reviews. 
Among them, graphical diagnosis methods are the most popular.
The mean residual life (MRL) plot \citep{davison1990models} 
uses the fact that, if $X$ follows a GPD, then for $v>u$, the 
MRL $E[X-v | X>v]$, if existing, is linear in $v$. 
The threshold is chosen to be the smallest $u$ such that the 
sample MRL is approximately linear above this point.
Parameter stability plots check whether the esimates of GPD 
parameters, possibly transformed to be comparable across 
different thresholds, are stable above some level of threshold.
\citet{drees2000make} suggested the Hill plot, which plots 
the Hill estimator of the shape parameter based on the top $k$
order statistics against $k$. Many variants of the Hill plot
have been proposed \citep[Section~4]{scarrott2012review}.
The threshold is the $k$th smallest order statistic beyond
which the parameter estimates are deemed to be stable.
The usual fit diagnostics such as quantile plots, return 
level plots, and probability plots can be helpful too, as
demonstrated in \citet{coles2001introduction}.
Graphical diagnostics give close inspection of the data,
but they are quite subjective and disagreement on a particular 
threshold can occur even among experts in the field; see, for 
example, a convincing demonstration of unclear choices using 
the Fort Collins precipitation data in \citet{scarrott2012review}.

% really six categories total -- graphical, probabilistic results, 
% computational, mixture models, goodness-of-fit, rules of thumb
Other selection methods can be grouped into various categories.
One is based on the asymptotic results about estimators of 
properties of the tail distribution. The threshold is selected by 
minimizing the asymptotic mean squared error (MSE) of the 
estimator of, for example, tail index \citep{beirlant1996}, 
tail probabilities \citep{hall1997estimation}, or
extreme quantiles \citep{ferreira2003optimising}. 
Theoretically sound as these methods are, their finite sample
properties are not well understood. Some require second order
assumptions, and computational (bootstrap) based estimators 
can require tuning parameters~\citep{Danielsson2001} or may 
not be satisfactory for small samples~\citep{ferreira2003optimising}. 
Irregardless, such resampling methods may be quite 
time-consuming in an analysis involving many locations.

A second category of methods are based on goodness-of-fit 
of the GPD, where the threshold is selected as the lowest level
above which the GPD provides adequate fit to the exceedances
\citep[e.g.,][]{davison1990models, dupuis1999, 
choulakian2001goodness, northrop2014improved}.
Goodness-of-fit tests are simple to understand and perform,
but the multiple testing issue for a sequence of tests in an ordered 
fashion have not been addressed to the best of our knowledge.
Methods in the third category are based on mixtures of a GPD for the 
tail and another distribution for the ``bulk'' joined at the threshold 
\citep[e.g.,][]{macdonald2011flexible, wadsworth2012likelihood}.
Treating the threshold as a parameter to estimate, these methods
can account for the uncertainty from threshold selection in inferences. 
However, there is little known about the asymptotic properties of 
this setup and how to ensure that the bulk and tail models are robust 
to one another in the case of misspecification.

Some automated procedures have been proposed. 
The simple naive method is \emph{a priori} or fixed threshold
selection based on expertise on the subject matter at hand. 
Various rules of thumb have been suggested; for example, 
select the top 10\% of the data
\citep[e.g.,][]{dumouchel1983estimating}, 
or the top square root of the sample size
\citep[e.g.,][]{ferreira2003optimising}.
Such one rule for all is not ideal in climate applications where high 
heterogeneity in data properties is the norm. The proportion of the number 
of rain days can be very different from wet tropical climates to dry 
subtropical climates; therefore the number of exceedance over the same 
time period can be very different across different climates. Additionally, 
the probability distribution of daily precipitation can also be different in 
different climates, affecting the speed the tails converge to the 
GPD~\citep{raoult2003rate}. 
Goodness-of-fit test based procedures can be 
automated to select the lowest one in a sequence of thresholds, at 
which the goodness-of-fittest is not 
rejected~\citep[e.g.,][]{choulakian2001goodness}. 
The error control, however, is challenging because of the ordered
nature of the hypotheses,  and the usual methods from multiple testing
such as false discovery rate
\citep[e.g.,][]{benjamini2010discovering, benjamini2010simultaneous}
cannot be directly applied.

% contributions
We propose an automated threshold selection procedure 
based on a sequence of goodness-of-fit tests with error 
control for ordered, multiple testing.
The very recently developed stopping rules for ordered 
hypotheses in \citet{g2015sequential} are adapted to control 
the familywise error rate (FWER), the probability of at least one type
I error in the whole family of tests \citep{shaffer1995multiple}, or
the false discovery rate (FDR),  the expected proportion of
incorrectly rejected null hypotheses among 
all rejections \citep{Benjamini1995, BY2001}.
They are applied to four goodness-of-fit tests at each candidate
threshold, including the Cram\'er--Von Mises test, 
Anderson--Darling test, Rao's score test, and Moran's test.
For the first two tests, the asymptotic null distribution of the
testing statistic is unwieldy \citep{choulakian2001goodness}.
Parametric bootstrap puts bounds on the approximate p-values
which would invalidate the stopping rules.
We propose a fast approximation based on the results of 
\citet{choulakian2001goodness} to facilitate the application.
The performance of the procedures are investigated in a
large scale simulation study, and recommendations are made.
The procedure is applied to annual maximum daily precipitation
return level mapping for three west coastal states of the US.
Interesting findings are revealed from different stopping rules.
The automated threshold selection procedure has applications 
in various fields, especially when batch processing of massive
datasets is needed.

% outline of paper
The outline of the paper is as follows. Section~\ref{s:seq_testing} 
presents the generalized Pareto model, its theoretical justification, 
and how to apply the automated sequential threshold testing procedure. 
Section~\ref{s:tests} introduces the tests proposed to be used in the 
automated testing procedure. A simulation study demonstrates the power 
of the tests for a fixed threshold under various misspecification 
settings and it is found that the Anderson--Darling test is most 
powerful in the vast majority of cases. A large scale simulation study in 
Section~\ref{s:sim} demonstrates the error control and performance 
of the stopping rules for multiple ordered hypotheses, both under 
the null GPD and a plausible alternative distribution. 
In Section~\ref{s:app}, we return to our motivating application and
derive return levels for extreme precipitation at hundreds of west
coastal US stations to demonstrate the usage of our method. 
A final discussion is given in Section~\ref{s:disc}.

\section{Automated Sequential Testing Procedure}
\label{s:seq_testing}

Threshold methods for extreme value analysis are based on
that, under general regularity conditions, the only possible
non-degenerate limiting distribution of properly rescaled 
exceedances of a threshold $u$ is the GPD as $u \to
\infty$~\citep[e.g.,][]{pickands1975statistical}.  
The GPD has cumulative distribution function
\begin{equation}
\label{eq:cdf}
F(y | \theta) = \begin{cases}
1 - \Big[1 + \frac{\xi y}{\sigma_u}\Big]^{-1/\xi} & \xi \neq 0, \quad
y > 0, \quad 1 + \frac{\xi y}{\sigma_u} > 0, \\
1 - \exp{\Big[-\frac{y}{\sigma_u}\Big]} & \xi = 0, \quad y > 0,
\end{cases}
\end{equation}
where $\theta = (\sigma_u, \xi)$, $\xi$ is a shape parameter, and
$\sigma_u > 0$ is a threshold-dependent scale parameter.
The GPD also has the property that for some threshold $v > u$, the
excesses follow a GPD with the same shape parameter, but a modified
scale $\sigma_v = \sigma_u + \xi(v - u)$.

Let $X_1, \ldots, X_n$ be a random sample of size $n$.
If $u$ is sufficiently high, the exceedances $Y_i = X_i - u$ for all 
$i$ such that $X_i > u$ are approximately a random sample from a GPD.
The question is to find the lowest threshold such that the GPD fits
the sample of exceedances over this threshold adequately.
Our solution is through a sequence of goodness-of-fit tests
\citep[e.g.,][]{choulakian2001goodness} or model specification tests
\citep[e.g.,][]{northrop2014improved} for the GPD to the exceedances
over each candidate threshold in an increasing order.
The multiple testing issues in this special ordered setting are
handled by the most recent stopping rules in \citet{g2015sequential}.

Consider a fixed set of candidate thresholds $u_1 < \ldots < u_l$.
For each threshold, there will be $n_i$ excesses, $i=1, \ldots, l$. 
The sequence of null hypotheses can be stated as
\begin{center}
$H_0^{(i)}$: The distribution of the $n_i$ exceedances above 
$u_i$ follows the GPD.
\end{center}
For a fixed $u_i$, many tests are available for this $H_0^{(i)}$.
An automated procedure can begin with $u_1$ and continue until 
some threshold $u_i$ provides an acceptance of $H_0^{(i)}$
\citep{choulakian2001goodness, thompson2009automated}.  
The problem, however, is that unless the test has high power, 
an acceptance may happen at a low threshold by chance and, 
thus, the data above the chosen threshold is contaminated. 
One could also begin at the threshold $u_l$ and descend until a
rejection occurs, but this would result in an increased type I error 
rate. The multiple testing problem obviously needs to be addressed, but 
the issue here is especially challenging because these tests are ordered; 
if $H_0^{(i)}$ is rejected, then $H_0^{(k)}$
will be rejected for all $1 \leq k < i$.
Despite the extensive literature on multiple testing and
the more recent developments on FDR control and its variants
\citep[e.g.,][]{Benjamini1995, BY2001, benjamini2010discovering,
  benjamini2010simultaneous},
no definitive procedure has been available for error control in 
ordered tests until the recent work of \citet{g2015sequential}.

We adapt the stopping rules of \citet{g2015sequential} to the
sequential testing of (ordered) null hypotheses $H_1, \ldots, H_l$.
Let $p_1, \ldots, p_l \in [0, 1]$ be the corresponding 
p-values of the $l$ hypotheses.
\citet{g2015sequential} transform the sequence of p-values to 
a monotone sequence and then apply the original method of
\citet{Benjamini1995} on the monotone sequence.
Two rejection rules are constructed, each of which returns a 
cutoff $\hat k$ such that $H_1, \ldots, H_{\hat k}$ are rejected. 
If no $\hat{k} \in \{1, \ldots, l\}$ exists, 
then no rejection is made. The first is called ForwardStop,
\begin{equation}
\label{eq:forwardstop}
\hat{k}_{\mathrm{F}} = \max \left\{k \in \{1, \ldots, l\}: -\frac{1}{k} \sum_{i=1}^k \log(1-p_i) \leq \alpha \right\},
\end{equation}
and the second is called StrongStop,
\begin{equation}
\label{eq:strongstop}
\hat{k}_{\mathrm{S}} = \max \left\{k \in \{1, \ldots, l\}:
  \exp\Big(\sum_{j=k}^l \frac{\log p_j}{j}\Big)   \leq \frac{\alpha
    k}{l} \right\},
\end{equation}
where $\alpha$ is a pre-specified level.

Under the assumption of independence among the tests, 
both rules were shown to control the FDR at level $\alpha$. 
In our setting, stopping at $k$ implies that goodness-of-fit 
of the GPD to the exceedances at the first $k$ thresholds 
$\{u_1, \ldots, u_k\}$ is rejected.
In other words, the first set of $k$ null hypotheses 
$\{H_1, \ldots, H_k\}$ are rejected. 
At each $H_0^{(i)}$, ForwardStop 
is a transformed average of the previous and current p-values, 
while StrongStop only accounts for the current and subsequent 
p-values. StrongStop provides an even stronger guarantee of error 
control; that is that the FWER is controlled at level $\alpha$. 
The tradeoff for stronger error control is reduced power to reject. 
Thus, the StrongStop rule tends to select a lower threshold than the 
ForwardStop rule. This is expected since higher thresholds are
more likely to approximate the GPD well, and thus provide higher
p-values. In this sense, ForwardStop could be thought of as more 
conservative (i.e., stopping at higher threshold by
rejecting more thresholds).

The stopping rules, combined with the sequential hypothesis 
testing, provide an automated selection procedure --- all 
that is needed are the levels of desired control for the 
ForwardStop and StrongStop procedures, and a set of thresholds. 
A caveat is that the p-values of the sequential tests here
are dependent, unlike the setup of \citet{g2015sequential}.
Nonetheless, the stopping rules may still provide some reasonable 
error control as their counter parts in the non-sequential multiple
testing scenario \citep{BY2001, blanchard2009adaptive}.
A simulation study is carried out in Section~\ref{s:sim} to assess 
the empirical properties of the two rules.

\section{The Tests}
\label{s:tests}

The automated procedure can be applied with any valid test
for each hypothesis $H_0^{(i)}$ corresponding to threshold $u_i$. 
Four existing goodness-of-fit tests that can be used are presented.
Because the stopping rules are based on transformed p-values, it is
desirable to have testing statistics whose p-values can be accurately
measured; bootstrap based tests that put a lower bound on the p-values 
(1 divided by the bootstrap sample size) may lead to premature stopping.
For the remainder of this section, the superscript $i$ is dropped.
We consider the goodness-of-fit of GPD to a sample of size $n$
of exceedances $Y = X - u$ above a fixed threshold $u$.

\subsection{Anderson--Darling and Cram\'{e}r--von Mises Tests} 
\label{ss:ad_cvm}

The Anderson--Darling and the Cram\'{e}r--von Mises tests for the 
GPD have been studied in detail \citep{choulakian2001goodness}.
Let $\hat\theta_n$ be the maximum likelihood estimator (MLE) of 
$\theta$ under $H_0$ from the the observed exceedances. 
Make the probability integral transformation based on $\hat\theta_n$
$z_{(i)} = F(y_{(i)} | \hat\theta_n)$, as in~\eqref{eq:cdf}, for the order
statistics of the exceedances $y_{(1)} < \ldots < y_{(n)}$. 
The Anderson--Darling statistic is 
\begin{equation*}
\label{eq:ad}
A_n^2 = -n - \frac{1}{n} \sum_{i=1}^n (2i - 1)\Big[\log(z_{(i)}) + \log(1 - z_{(n+1-i)}) \Big].
\end{equation*}
The Cram\'{e}r--von Mises statistic is
\begin{equation*}
\label{eq:cvm}
W_n^2 = \sum_{i=1}^n \big[z_{(i)} - \frac{2i - 1}{2n}\big]^2 + \frac{1}{12n}.
\end{equation*}

The asymptotic distributions of $A_n^2$ and $W_n^2$ are unwieldy,
both being sum of weighted chi-squared variables with one degree
of freedom with weight found from the eigenvalues of an integral
equation \citep[Section~6]{choulakian2001goodness}.
The distributions depend only on the estimate of $\xi$.
The tests are often applied by referring to a table of a few upper
tail percentiles of the asympotic distributions
\citep[Table~2]{choulakian2001goodness}, or through bootstrap.
In either case, the p-values are truncated by a lower bound.
Such truncation of a smaller p-value to a larger one can be proven to
weaken the stopping rules given in~\eqref{eq:forwardstop} 
and~\eqref{eq:strongstop}. In order to apply these tests in the 
automated sequential setting, more accurate p-values for the tests 
are needed.

We provide two remedies to the table in
\citet{choulakian2001goodness}.
First, for $\xi$ values in the range of $(-0.5, 1)$, which is
applicable for most applications, we enlarge the table to a much 
finer resolution through a pre-set Monte Carlo method.
For each $\xi$ value from $-0.5$ to $1$ incremented by $0.1$, 
2,000,000 replicates of $A_n^2$ and $W_n^2$ are generated with
sample size $n = 1,000$ to approximate their asymptotic distributions.
A grid of upper percentiles from 0.999 to 0.001 for each $\xi$ value
is produced and saved in a table for future fast reference.
Therefore, if $\hat\xi_n$ and the test statistic falls in the range
of the table, the p-value is computed via log-linear interpolation.

The second remedy is for observed test statistics that are greater
than that found in the table (implied p-value less than 0.001).
As Choulakian pointed out (in a personal communication),
the tails of the asymptotic distributions are exponential, which 
can be confirmed using the available tail values in the table. 
For a given $\hat\xi$, regressing $-\log(\text{p-value})$ on the
upper tail percentiles in the table, for example, from 0.05 to
0.001, gives a linear model that can be extrapolated to approximate
the p-value of statistics outside of the range of the table.
This approximation of extremely small p-values help reduce loss of
power in the stopping rules.

The two remedies make the two tests very fast and are 
applicable for most applications with $\xi \in (-0.5, 1)$. 
For $\xi$ values outside of $(-0.5, 1)$, although slow, one can 
use parametric  bootstrap to obtain the distribution of the test
statistic, understanding that the p-value has a lower bound.
The methods are implemented in R package \texttt{eva} \citep{Rpkg:eva}.

\subsection{Moran's Test}
\label{ss:moran}
Moran's goodness-of-fit test is a byproduct of the maximum product
spacing (MPS)  estimation for estimating the GPD parameters. 
MPS is a general method that allows efficient parameter estimation in
non-regular cases where the MLE fails or the
information matrix does not exist \citep{cheng1983estimating}.
It is based on the fact that if the exceedances are indeed from the
hypothesized distribution, their probability integral transformation would
behave like a random sample from the standard uniform distribution.
Consider the ordered exceedances $y_{(1)}  < \ldots < y_{(n)}$.
Define the spacings as 
\begin{equation*}
D_{i}(\theta) = F(y_{(i)} | \theta) - F(y_{(i-1)} | \theta)
\end{equation*}
for $i=1, 2, \ldots, n+1$ with $F(y_{(0)} | \theta) \equiv 0$ 
and $F(y_{(n+1)} | \theta) \equiv 1$. 
The MPS estimators then found by minimizing
\begin{equation*}
\label{eq:morans}
M(\theta) = - \sum\limits_{i=1}^{n+1} \log D_{i}(\theta).
\end{equation*}
As demonstrated in \citet{wong2006note}, the MPS method is especially
useful for GPD estimation in the non-regular cases of \citet{smith1985maximum}.
In cases where the MLE exists, the MPS estimator may
have an advantage of being numerical more stable for small samples,
and have the same properties as the MLE asymptotically.

The objective function evaluated at the MPS estimator 
$\check\theta$ is Moran's statistic \citep{moran1953random}.
\citet{cheng1989goodness} showed that, under the null hypothesis,
Moran's statistic is normally  distributed and when properly center
and scaled, has an asymptotic  chi-square approximation. 
Define
\begin{align*}
& \mu_M = (n+1)\big(\log(n+1) + \gamma\big) - \frac{1}{2} - \frac{1}{12(n+1)}, \\
& \sigma_M^2 = (n+1)\left(\frac{\pi^2}{6} - 1\right) - \frac{1}{2} - \frac{1}{6(n+1)},
\end{align*}
where $\gamma$ is Euler's constant. 
Moran's goodness-of-fit test statistic is
\begin{equation*}
T(\check{\theta}) = \frac{M(\check{\theta}) + 1 - C_1}{C_2},
\end{equation*}
where $C_1 = \mu_M - (n/ 2)^{\frac{1}{2}} \sigma_M$ and
$C_2 = (2n)^{- \frac{1}{2}} \sigma_M$.
Under the null hypothesis asymptotically, $T(\check{\theta})$ 
follows a chi-square distribution with $n$ degrees of freedom. 
\citet{wong2006note} show that for GPD data, the test empirically
holds its size for samples as small as ten.

\subsection{Rao's Score Test}
\label{ss:score}

\citet{northrop2014improved} considered a piecewise constant 
specification of the shape parameter as the alternative hypothesis.
For a fixed threshold $u$, a set of $k$ higher thresholds are
specified to  generate intervals on the support space. 
That is, for the set of thresholds $u_0 = u < u_1 < \ldots < u_k$,
the shape parameter is given a piecewise representation
\begin{equation}
\label{eq:piecewise}
\xi(x) = \begin{cases}
\xi_i & u_i < x \leq u_{i+1} \quad  i=0, \ldots, k-1,  \\
\xi_k & x > u_k.
\end{cases}
\end{equation}
The null hypothesis is tested as 
$H_0: \xi_0 = \xi_1 = \cdots =\xi_k$. 
Rao's score test has the advantage that only restricted MLE
$\tilde\theta$ under $H_0$ is needed, in contrast to the
asymptotically equivalent likelihood ratio test or Wald test.
The testing statistic is
\[
S(\tilde{\theta}) = 
U(\tilde{\theta})^T I^{-1}(\tilde{\theta}) U(\tilde{\theta}),
\]
where $U$ is the score function and $I$ is the fisher information
matrix, both evaluated at the restricted MLE $\tilde\theta$.
Given that $\xi > -0.5$~\citep{smith1985maximum}, the asymptotic null 
distribution of $S$ is $\chi^2_{k}$.

\citet{northrop2014improved} tested suitable thresholds in an ascending
manner, increasing the initial threshold $u$ and continuing the 
testing of $H_0$. They suggested two possibilities for automation. 
First, stop testing as soon as an acceptance occurs, but the p-values
are not guaranteed to be non-decreasing for higher starting thresholds. 
Second, stop as soon as all p-values for testing at
subsequent higher thresholds are above a certain significance level. 
The error control under multiple, ordered testing were not addressed.

\subsection{A Power Study}
\label{ss:power}

The power of the four goodness-of-fit tests are examined 
in an individual, non-sequential testing framework.
The data generating schemes in \citet{choulakian2001goodness} 
are used, some of which are very difficult to distinguish 
from the GPD:
\begin{itemize}
\item
Gamma with shape 2 and scale 1.

\item
Standard lognormal distribution (mean 0 and scale 1 on log scale).

\item
Weibull with scale 1 and shape 0.75.

\item
Weibull with scale 1 and shape 1.25.

\item
50/50 mixture of GPD($1, -0.4$) and GPD($1, 0.4$).

\item
50/50 mixture of GPD($1, 0$) and GPD($1, 0.4$).

\item
50/50 mixture of GPD($1, -0.25$) and GPD($1, 0.25$).
\end{itemize}
Finally, the GPD($1, 0.25$) was also used to check the sizes.
Four sample sizes were considered: 50, 100, 200, 400.
For each scenario, 10,000 samples are generated.
The four tests were applied to each sample, with a 
rejection recorded if the p-value is below 0.05. 
For the score test, a set of thresholds were set according 
to the deciles of the generated data.

\begin{table}[htbp]
  \centering
  \caption{Empirical rejection rates of four goodness-of-fit tests for
    GPD under various data generation schemes with nominal size 0.05.
    GPDMix(a, b) refers to a 50/50 mixture of GPD(1, a) and GPD(1, b). 
}
    \begin{tabular}{l rrrr rrrr}
    \toprule
    Sample Size & \multicolumn{4}{c}{50} & \multicolumn{4}{c}{100} \\
    \cmidrule(lr){2-5}\cmidrule(lr){6-9}
    Test & Score & Moran & AD & CVM  & Score & Moran & AD & CVM \\
    \midrule
    Gamma(2, 1) & 7.6   & 9.7   & 47.4  & 43.5  & 8.0   & 14.3  & 64.7  & 59.7 \\
    LogNormal & 6.0   & 5.9   & 13.3  & 8.6   & 5.4   & 8.2   & 28.3  & 23.4 \\
    Weibull(0.75) & 11.5  & 7.8   & 55.1  & 23.5  & 12.1  & 9.5   & 65.1  & 39.4 \\
    Weibull(1.25) & 6.6   & 11.3  & 29.1  & 27.3  & 5.6   & 12.5  & 20.8  & 19.2 \\
    GPDMix($-$0.4, 0.4) & 11.4  & 7.5   & 19.2  & 9.9   & 16.0  & 8.6   & 24.3  & 20.4 \\
    GPDMix(0, 0.4) & 7.8   & 6.0   & 6.5  & 5.9   & 7.4   & 5.9   & 9.6   & 5.6 \\
    GPDMix($-$0.25, 0.25) & 8.1   & 6.5   & 6.0  & 7.4   & 8.9   & 6.5   & 11.1  & 8.4 \\
    GPD(1, 0.25) & 6.9   & 5.5   & 6.7   & 6.1   & 5.0   & 5.2   & 5.2   & 5.2 \\
    \midrule
    Sample Size & \multicolumn{4}{c}{200} & \multicolumn{4}{c}{400} \\
    \cmidrule(lr){2-5}\cmidrule(lr){6-9}
    Test & Score & Moran & AD & CVM  & Score & Moran & AD & CVM \\
    \midrule
    Gamma(2, 1) & 15.4  & 23.3  & 95.3  & 93.1  & 36.5  & 42.2  & 100.0 & 100.0 \\
    LogNormal & 5.7   & 11.9  & 69.3  & 59.7  & 7.9   & 19.1  & 97.8  & 95.0 \\
    Weibull(0.75) & 16.5  & 10.7  & 84.8  & 66.4  & 32.1  & 14.6  & 98.2  & 93.0 \\
    Weibull(1.25) & 8.0   & 14.7  & 40.9  & 36.7  & 15.5  & 19.2  & 79.8  & 74.6 \\
    GPDMix($-$0.4, 0.4) & 31.5  & 9.7   & 45.1  & 44.0  & 63.8  & 11.9  & 79.9  & 80.2 \\
    GPDMix(0, 0.4) & 8.9   & 6.5   & 8.8   & 7.3   & 12.3  & 6.2   & 10.8  & 10.3 \\
    GPDMix($-$0.25, 0.25) & 13.9  & 6.7   & 16.6  & 14.8  & 26.2  & 7.9   & 33.0  & 32.4 \\
    GPD(1, 0.25) & 5.3   & 5.8   & 7.2   & 5.2   & 4.7   & 5.3   & 5.8   & 4.7 \\
    \bottomrule
    \end{tabular}
  \label{tab:powerstudy}
\end{table}

The rejection rates are summarized in Table~\ref{tab:powerstudy}. 
Samples in which the MLE failed were removed, 
which accounts for roughly 10.8\% of the Weibull samples with 
shape 1.25 and sample size 400, and around 10.7\% for the 
Gamma distribution with sample size 400. Decreasing the sample 
size in these cases actually decreases the percentage of failed 
MLE samples. This may be due to the shape of these two 
distributions, which progressively become more distinct from the 
GPD as their shape parameters increase. In the other distribution 
cases, no setting resulted in more than a 0.3\% failure rate.
As expected, all tests appear to hold their sizes, and 
their powers all increase with sample size.
The mixture of two GPDs is the hardest to detect. 
For the GPD mixture of shape parameters 0 and 0.4, quantile matching 
between a single large sample of generated data and the fitted GP 
distribution shows a high degree of similarity. 
In the vast majority of cases, the Anderson--Darling test appears to
have the highest power, followed by the Cram\'er--von Mises test.
Between the two, the Anderson--Darling statistic is a modification of
the Cram\'er--von Mises statistic giving more weight to observations
in the tail of the distribution, which explains the edge of the former.

\section{Simulation Study of the Automated Procedures}
\label{s:sim}

\begin{figure}[!ht]
   \centering
      \includegraphics[width=\textwidth]{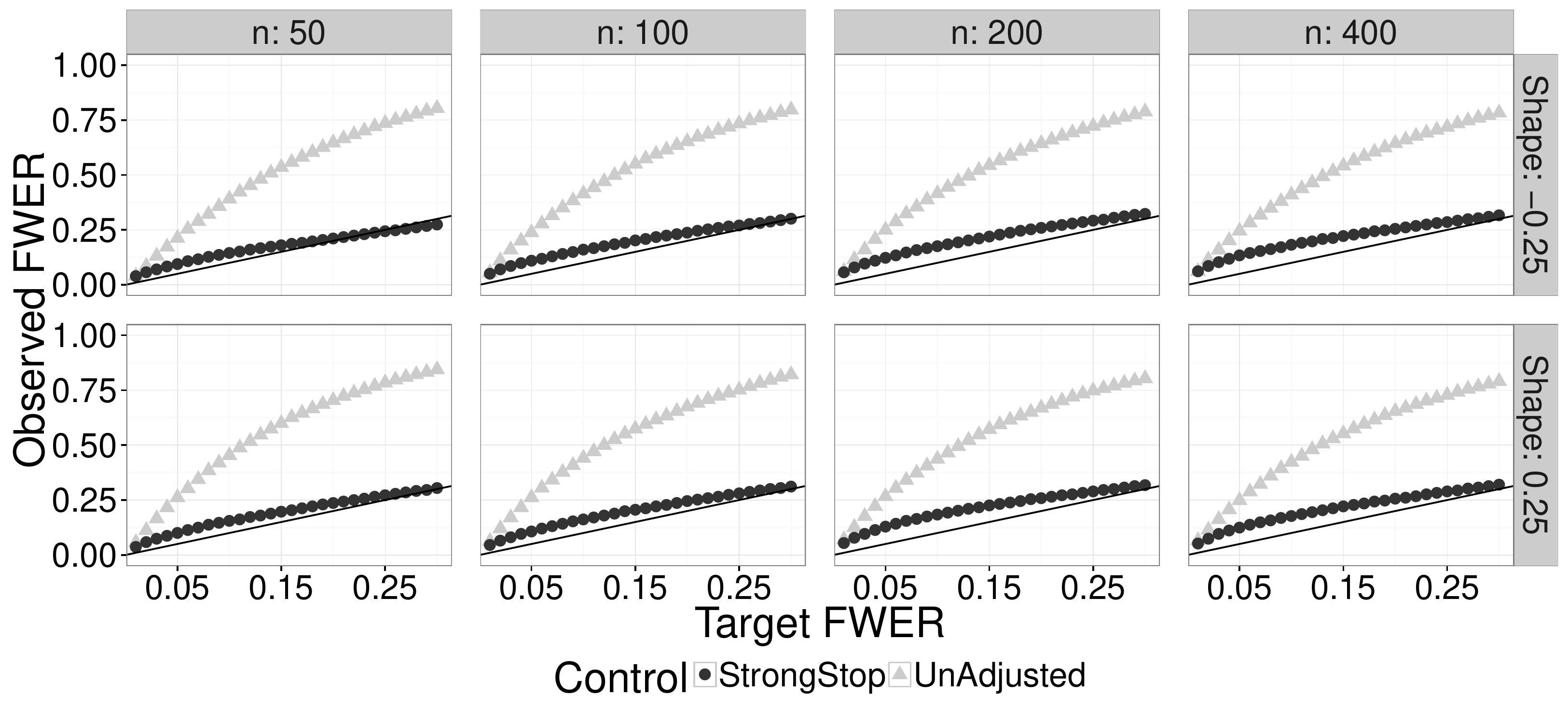}
    \caption{Observed FWER for the Anderson--Darling test (using StrongStop and 
    no adjustment) versus expected FWER at various nominal levels. The 45 
    degree line indicates agreement between the observed and expected rates 
    under H0.}
    \label{fig:FWER_Check}
\end{figure}

The empirical properties of the two stopping rules for the Anderson--Darling 
test are investigated in simulation studies. To check the empirical 
FWER of the StrongStop rule, data only need to be generated under 
the null hypothesis. For $n \in \{50, 100, 200, 400\}$, $\xi \in \{-0.25, 
0.25\}$, $\mu=0$, and $\sigma=1$, 10,000 GPD samples were generated in 
each setting of these parameters. Ten thresholds are tested by locating 
ten percentiles, 5 to 50 by 5. Since the data is generated 
from the GPD, data above each threshold is also distributed as GP, 
with an adjusted scale parameter. Using the StrongStop procedure and 
with no adjustment, the observed FWER is compared to the expected 
rates for each setting at various nominal levels. At a given 
nominal level and setting of the parameters, the observed FWER is 
calculated as the number of samples with a rejection of $H_0$ at any 
of the thresholds, divided by the total number of samples. The results 
of this study can been seen in Figure~\ref{fig:FWER_Check}.

It can be seen that the observed FWER under $H_0$ using the StrongStop 
procedure is nearly in agreement with the expected rate at most nominal 
levels for the Anderson--Darling test (observed rate is always 
within 5\% of the expected error rate). 
However, using the naive, unadjusted 
stopping rule (simply rejecting for any p-value less than the nominal 
level), the observed FWER is generally 2--3 times the expected rate 
for all sample sizes.

\begin{figure*}[tbp]
    \centering
      \includegraphics[width=\textwidth]{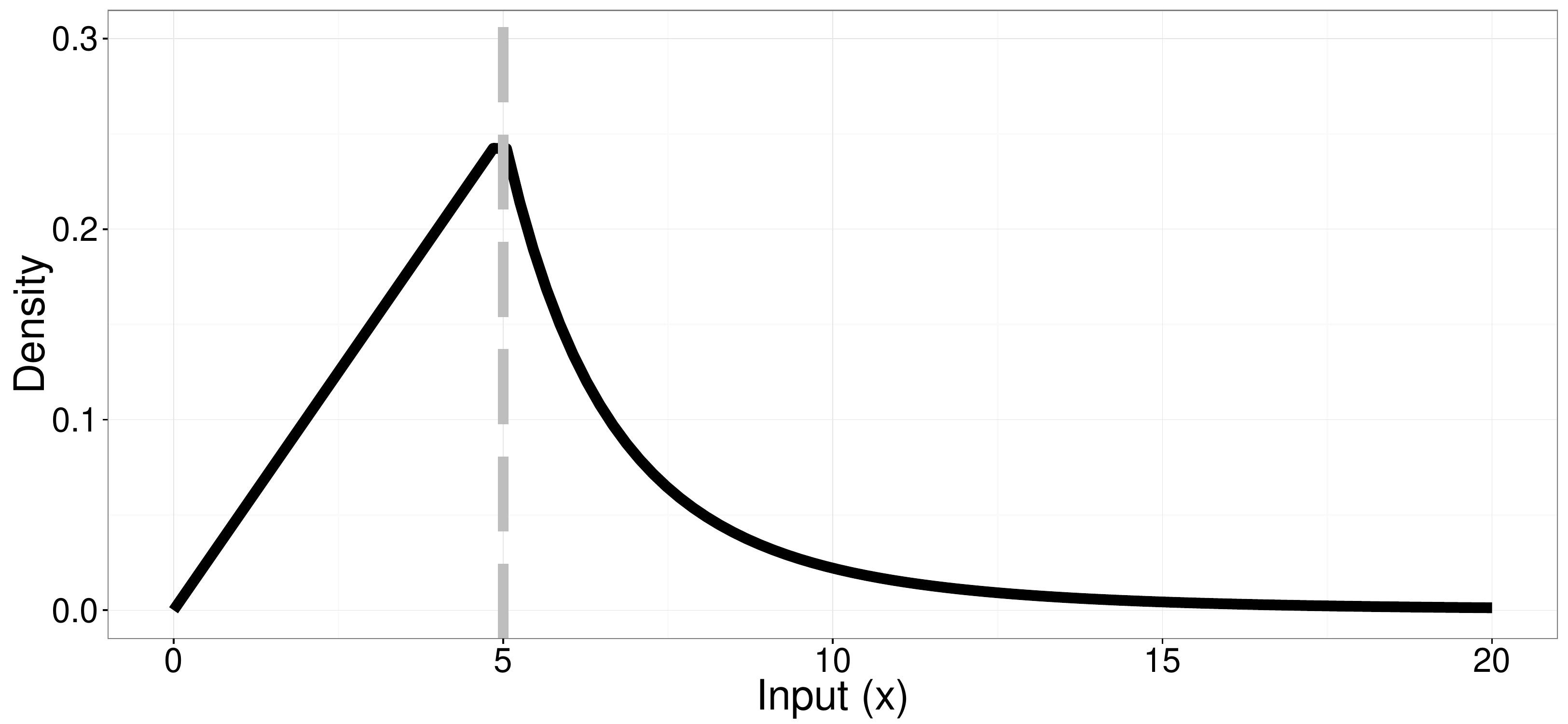}
    \caption{Plot of the (scaled) density of the mixture distribution used 
    to simulate misspecification of $H_0$. Vertical line indicates the 
    continuity point of the two underlying distributions.}
    \label{fig:MixtureDistribution}
\end{figure*}

It is of interest to investigate the performance of the ForwardStop and 
StrongStop in selecting a threshold under misspecification. To check the 
ability of ForwardStop and StrongStop to control the false discover 
rate (FDR), data need to be generated under misspecification. Consider 
the situation where data is generated from a 50/50 mixture of 
distributions. Data between zero and five are generated from a Beta 
distribution with $\alpha = 2$, $\beta=1$ and scaled such that the 
support is on zero to five. Data above five is generated from the 
GPD($\mu=5$, $\sigma=2$, $\xi=0.25$). 
Choosing misspecification in this 
way ensures that the mixture distribution is at least continuous at 
the meeting point. See Figure~\ref{fig:MixtureDistribution} for 
a visual assessment.

A total of $n=1000$ observations are generated, with $n_1 = 500$ 
from the Beta distribution and $n_2 = 500$ from the GP distribution. 
1000 datasets are simulated and 50 thresholds are tested, starting 
by using all $n=1000$ observations and removing the 15 lowest 
observations each time until the last threshold uses just 250 
observations. In this way, the correct threshold is given when 
the lowest $n_1$ observations are removed.

The results are presented here. The correct threshold is the 34th, 
since $n_1 = 500$ and $\ceil{500 / 15} = 34$. Rejection of less than 
33 tests is problematic  since it allows contaminated data to 
be accepted. Thus, a conservative selection criteria is desirable. 
Rejection of more than 33 tests is okay, although some non-contaminated 
data is being thrown away. By its nature, the StrongStop procedure has 
less power-to-reject than ForwardStop, since it has a stricter 
error control (FWER versus FDR).

\begin{figure*}[tbp]
    \centering
      \includegraphics[width=\textwidth]{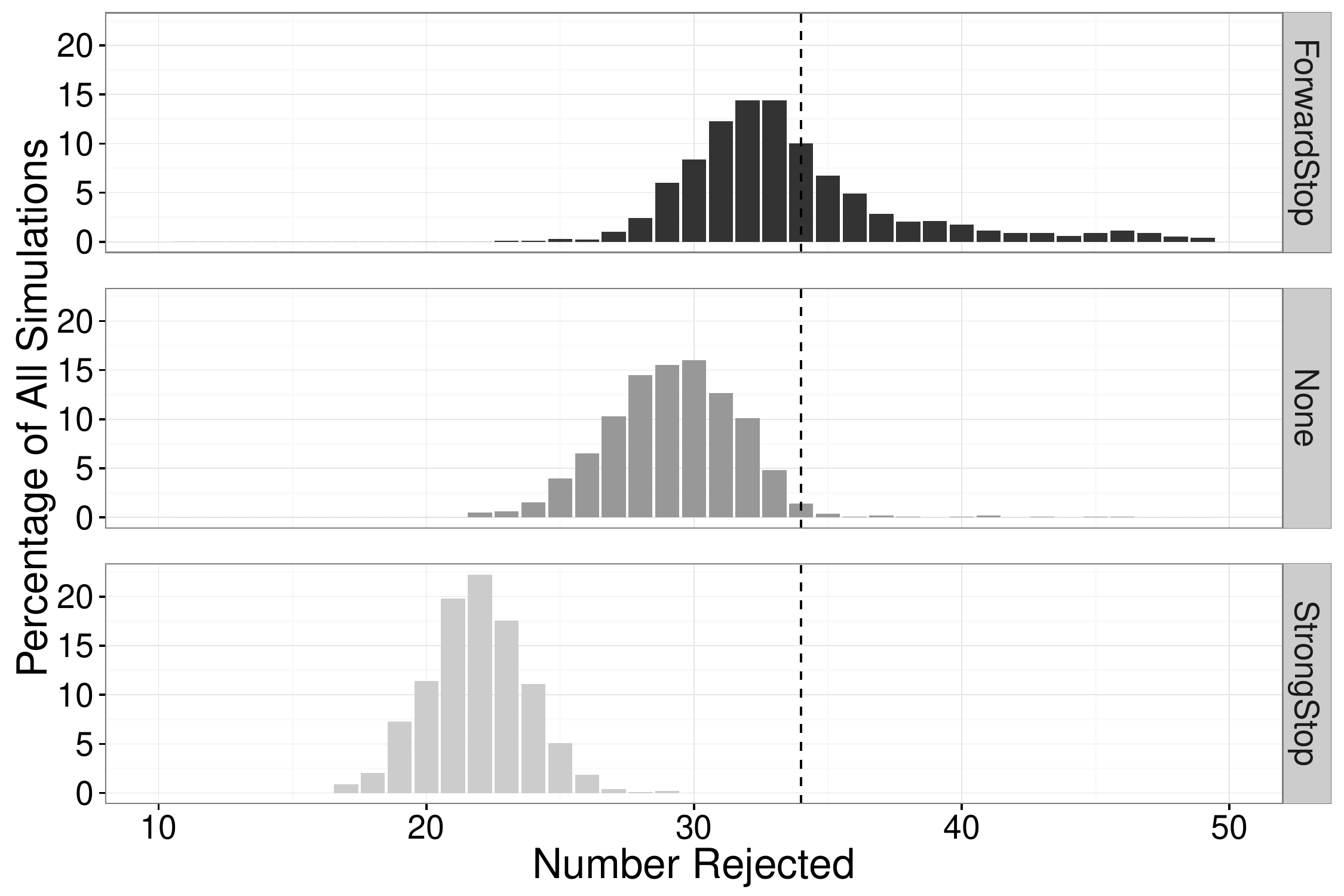}
    \caption{Percent frequency of the number of rejections for the 
    Anderson--Darling test and different error control procedures, at the 
    5\% nominal level. The correct number of rejections is 34. This is 
    for the simulation setting under sequential misspecification described 
    in Section~\ref{s:sim}.}
    \label{fig:Threshold_Sliced}
\end{figure*}

Figure~\ref{fig:Threshold_Sliced} displays the frequency distribution 
for threshold choice using the Anderson--Darling test at the 
5\% nominal level for the three stopping rules.
There is a clear hierarchy in terms of power-to-reject between 
the ForwardStop, StrongStop, and no adjustment procedures. 
StrongStop, as expected, provides the least power-to-reject, 
with all 1000 simulations selecting a threshold below the correct 
one. ForwardStop is more powerful than the no adjustment 
procedure and on average selects a higher threshold. The median 
number of thresholds rejected is 33, 29, and 22 for ForwardStop, 
no adjustment, and StrongStop respectively.

\begin{figure*}[tbp]
    \centering
      \includegraphics[scale = 0.4]{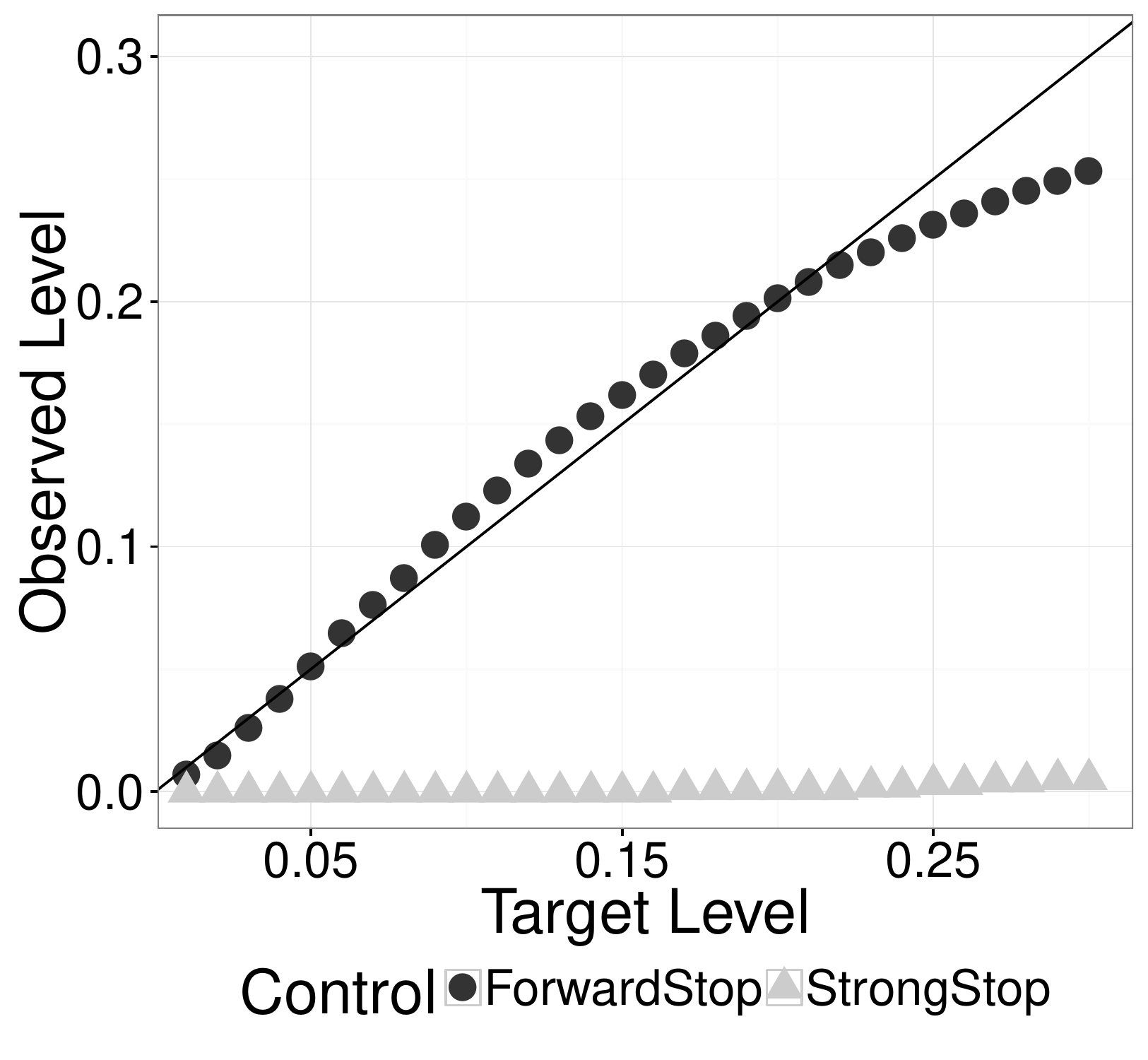}
    \caption{Observed FDR (using ForwardStop) and observed FWER 
    (using StrongStop) versus expected FDR and FWER respectively 
    using the Anderson--Darling test, at various nominal levels. This 
    is for the sequential simulation setting under misspecification 
    described in Section~\ref{s:sim}. The 45 degree line indicates agreement 
    between the observed and expected rates.}
    \label{fig:FWER_FDR_Combined}
\end{figure*}

The observed versus expected FDR using ForwardStop and the observed 
versus expected FWER using StrongStop using the Anderson--Darling test 
for the data generated under misspecification can be seen in 
Figure~\ref{fig:FWER_FDR_Combined}. There appears to be reasonable 
agreement between the expected and observed FDR rates using ForwardStop, 
while StrongStop has observed FWER rates well below the expected rates.

% performance evaluation of tests and stopping rules
To further evaluate the performance of the combination of tests 
and stopping rules, it is of interest to know how well the data 
chosen above each threshold can estimate parameters of interest. 
Two such parameters are the shape and return level. The $N$ year 
return level~\citep[e.g.,][Section~4.3.3]{coles2001introduction} 
is given by 
\begin{equation}
z_N = 
\begin{cases} 
  u + \frac{\sigma}{\xi}[(N n_y \zeta_u)^\xi - 1], & \xi \neq 0, \\
  u + \sigma \log(N n_y \zeta_u), & \xi = 0.
\end{cases}
\end{equation}
for a given threshold $u$, where $n_y$ is the number of observations 
per year, and $\zeta_u$ is the rate, or proportion of the data 
exceeding $u$. The rate parameter has a natural estimator simply 
given by the number of exceeding observations divided by the total 
number of observations. A confidence interval for $z_N$ can easily 
be found using the delta method, or preferably and used here, profile 
likelihood. The `true' return levels are found by letting $u = 34$, 
$n_y = 365$, and treating the data as the tail of some larger dataset, 
which allows computation of the rate $\zeta_u$ for each threshold. 

\begin{figure*}[tbp]
    \centering
      \includegraphics[width=\textwidth]{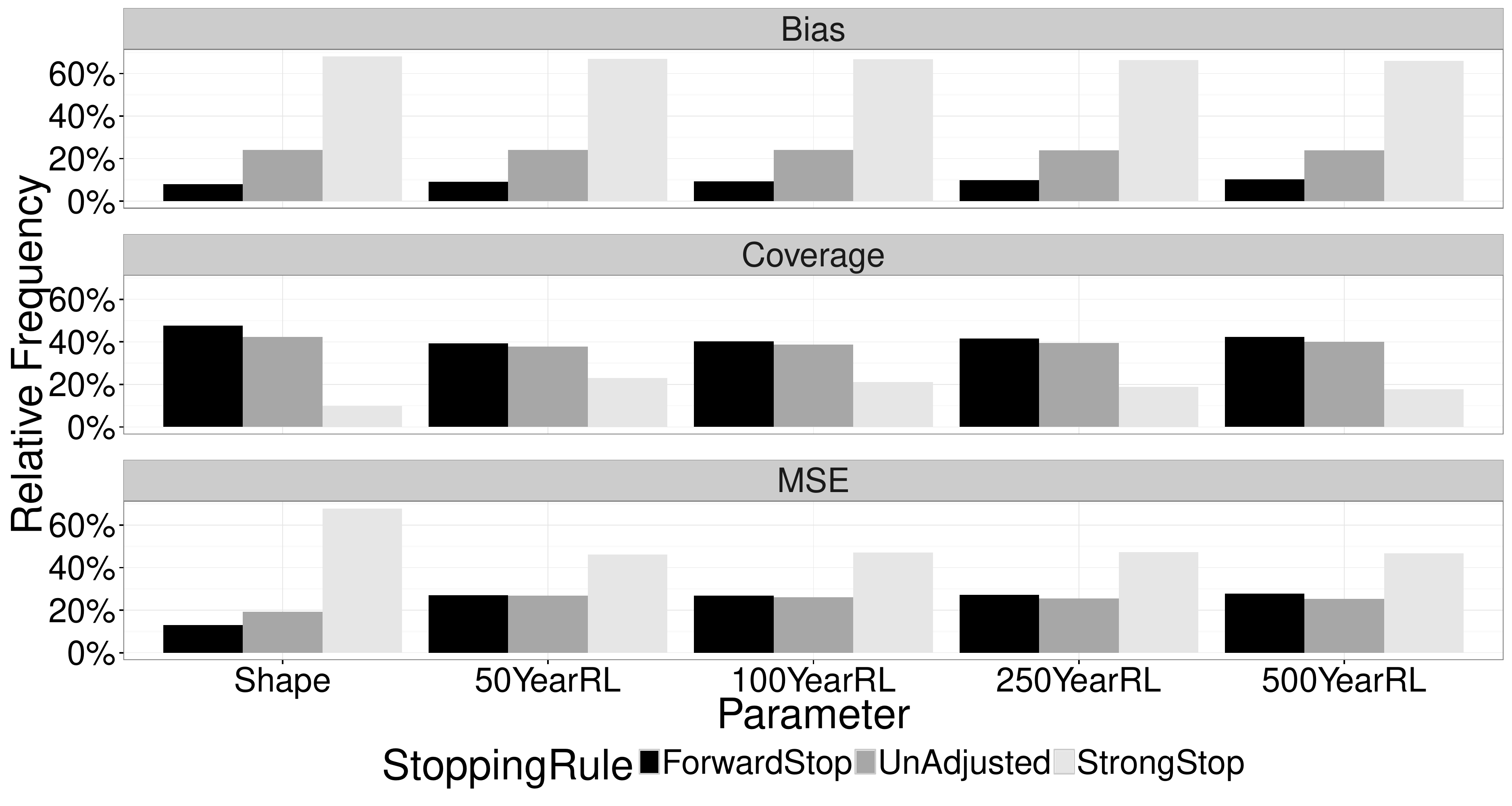}
    \caption{Average performance comparison of the three stopping rules 
    in the simulation study under misspecification, 
    using the Anderson--Darling test for various parameters. Shown are the 
    relative frequencies of the average value of each metric (bias, squared 
    error, and coverage) for each stopping rule and parameter of interest.}
    \label{fig:StoppingRuleCompare}
\end{figure*}

For ease of presentation, focus will be on the performance of the 
Anderson--Darling test in conjunction with the three stopping 
rules. In each of the 1000 simulated datasets 
(seen in Figure~\ref{fig:MixtureDistribution}), the threshold 
selected for each of the three stopping rules is used to determine 
the bias, squared error, and confidence interval coverage 
(binary true/false) for the shape parameter and 50, 100, 250, and 
500 year return levels. An average of the bias, squared error, 
and coverage across the 1000 simulations is taken, 
for each stopping rule and parameter value. For each parameter and 
statistic of interest (mean bias, mean squared error (MSE), and mean coverage), 
a relative percentage is calculated for the three stopping rules. A 
visual assessment of this analysis is provided in 
Figure~\ref{fig:StoppingRuleCompare}.

It is clear from the result here that ForwardStop is the 
most preferable stopping rule in application. For all parameters, 
it has the smallest average bias (by a proportion of 2-1) and 
highest coverage rate. In addition, the MSE is comparable or 
smaller than the unadjusted procedure in all cases. Arguably, 
StrongStop has the worst performance, obtaining the highest 
average bias and MSE, and the lowest coverage rates for all 
parameters. Replacing the Anderson--Darling test with the other 
tests provides similar results (not shown here).

\section{Return Level Mapping of Extreme Precipitation}
\label{s:app}

One particularly useful application of the automated threshold 
selection method is generating an accurate return level map of 
extreme precipitation in the three western US coastal states 
of California, Oregon, and Washington. The automated procedure 
described in Section~\ref{s:seq_testing} provides a method to 
quickly obtain an accurate map without the need for visual 
diagnostics at each site. Return level maps are of great 
interest to hydrologists~\citep{katz2002statistics} and 
can be used for risk management and land 
planning~\citep{blanchet2010mapping, lateltin1999hazard}.

Daily precipitation data is available for tens of thousands 
of surface sites around the world via the Global Historical 
Climatology Network (GHCN). A description of the data network 
can be found in~\citet{menne2012overview}. After an initial 
screening to remove sites with less than 50 years of 
available data, there are 720 remaining sites across the 
three chosen coastal states. 

As the annual maximum daily amount of precipitation mainly 
occurs in winter, only the winter season (November - March) 
observations are used in modeling. A set of thresholds for each 
site are chosen based on the data percentiles; for each site the 
set of thresholds is generated by taking the 75th to 97th 
percentiles in increments of 2, then in increments of 0.1 
from the 97th to 99.5th percentile. This results in 37 
thresholds to test at each site. If consecutive percentiles 
result in the same threshold due to ties in data, 
only one is used to guarantee uniqueness of thresholds and 
thus reducing the total number of thresholds tested at that site.

As discussed in the beginning of Section~\ref{s:seq_testing}, 
modeling the exceedances above a threshold with generalized 
Pareto requires the exceedances to be independent. This is 
not always the case with daily precipitation data; a large 
and persistent synoptic system can bring large storms 
successively. The extremal index is a measure of the 
clustering of the underlying process at extreme levels. Roughly 
speaking, it is equal to (limiting mean cluster size)$^{-1}$. 
It can take values from 0 to 1, with independent series 
exhibiting a value of exactly 1. To get a sense for the 
properties of series in this dataset, the extremal index 
is calculated for each site using the 75th percentile as the 
threshold via the R package \texttt{texmex}~\citep{Southworth2013}. 
In summary, the median estimated extremal index for all 720 
sites is 0.9905 and 97\% of the sites have an extremal index 
above 0.9. Thus, we do not do any declustering on the data. 
A more thorough description of this process can be found in 
\citet{Ferro2003} and \citet{heffernan2012extreme}.

\begin{figure*}[tbp]
    \centering
      \includegraphics[width=\textwidth]{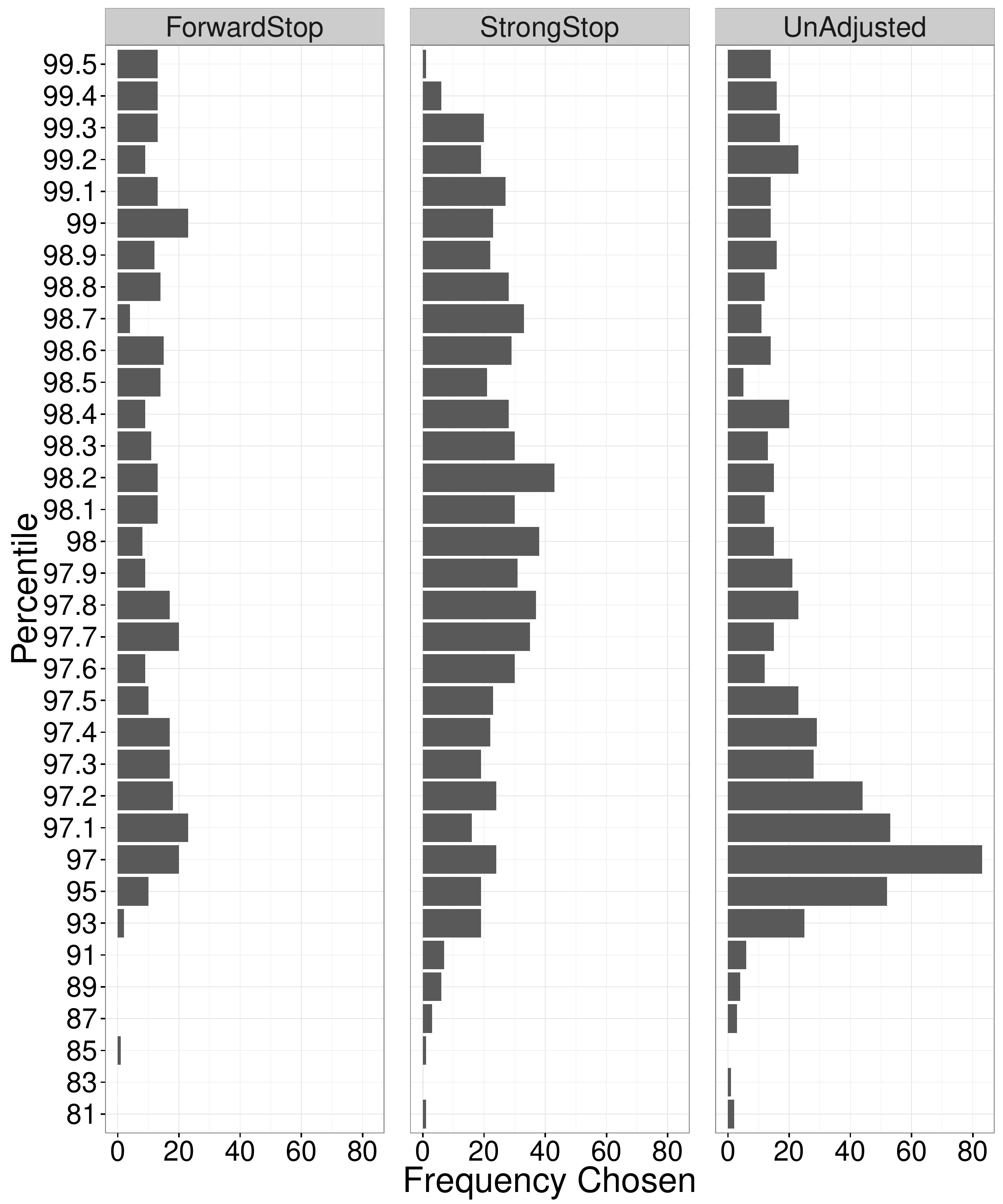}
    \caption{Distribution of chosen percentiles (thresholds) for the 720 
    western US coastal sites, as selected by each stopping rule. Note that 
    this does not include sites where all thresholds were rejected by the 
    stopping rule.}
    \label{fig:ChosenThresholds}
\end{figure*}

The Anderson--Darling test  is used to test the set of thresholds 
at each of the 720 sites, following the procedure outlined in 
Section~\ref{ss:ad_cvm}. This is arguably the most powerful test 
out of the four examined in Section~\ref{ss:power}. Three 
stopping rules are used --- ForwardStop, StrongStop, and with no 
adjustment, which proceeds in an ascending manner until an acceptance 
occurs. Figure~\ref{fig:ChosenThresholds} shows the distribution 
of chosen percentiles for the 720 sites using each of the three 
stopping rules.

\begin{figure}[!ht]
   \centering
      \includegraphics[width=\textwidth]{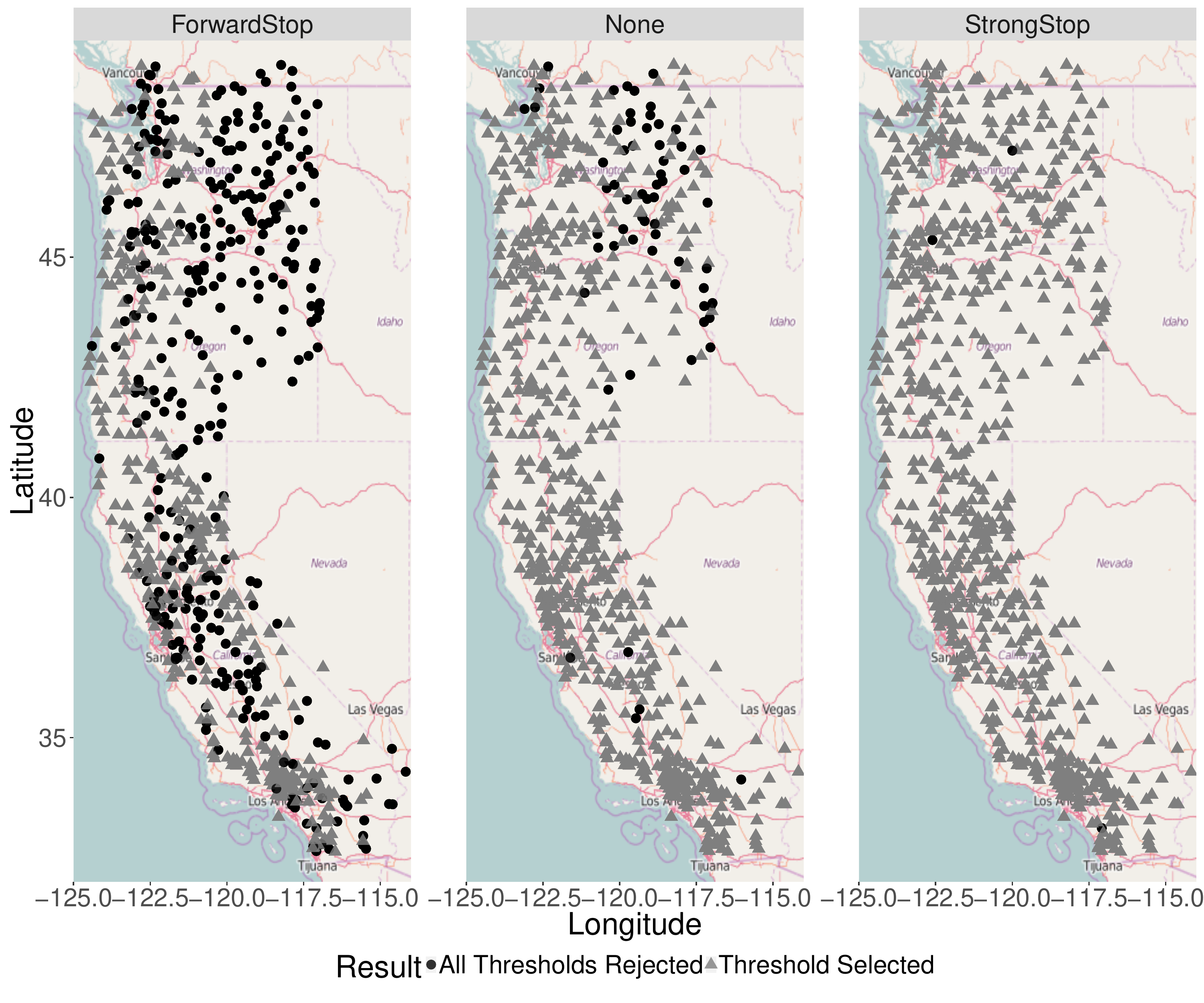}
    \caption{Map of US west coast sites for which all thresholds were 
    rejected (black) and for which a threshold was selected (grey), by 
    stopping rule.}
    \label{fig:All_Thresholds_Rejected}
\end{figure}

As expected, ForwardStop is the most conservative, rejecting all 
thresholds at 348 sites, with the unadjusted procedure rejecting 63, 
and StrongStop only rejecting all thresholds at 3 sites. 
Figure~\ref{fig:All_Thresholds_Rejected} shows the geographic 
representation of sites in which all thresholds are rejected. 
Note that there is a pattern of rejections by ForwardStop, 
particularly in the eastern portion of Washington and Oregon,
and the Great Valley of California. This may be attributed to 
the climate differences in these regions -- rejected sites had 
a smaller number of average rain days than non-rejected sites 
(30 vs. 34), as well as a smaller magnitude of precipitation (an 
average of 0.62cm vs. 1.29cm). The highly selective feature of 
the ForwardStop procedure is desired as it suggests not to fit 
GPD at even the highest threshold at these sites, a guard that 
is not available from those unconditionally applied, one-for-all 
rules.

\begin{figure}[!ht]
   \centering
      \includegraphics[width=\textwidth]{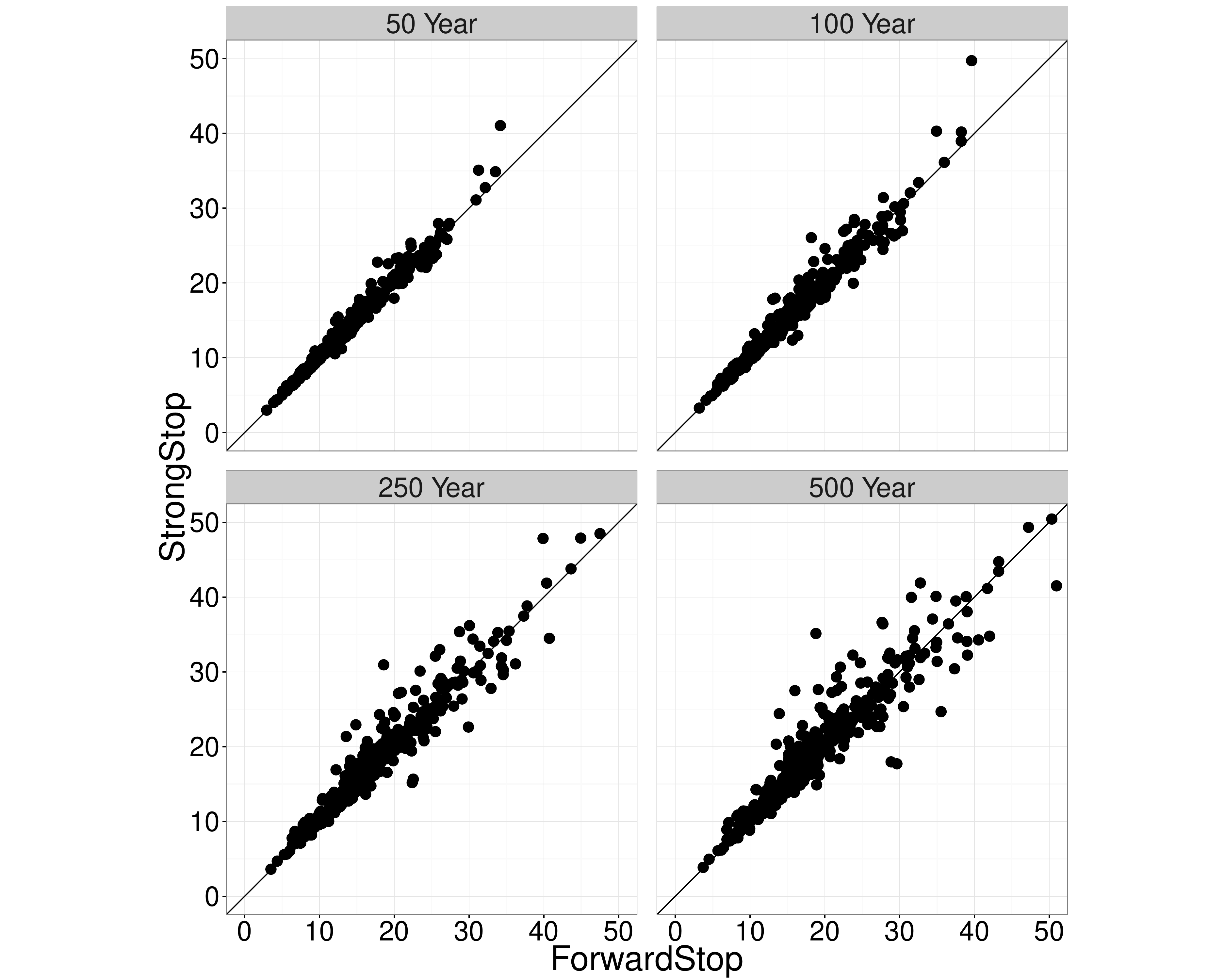}
    \caption{Comparison of return level estimates based on the chosen 
    threshold for ForwardStop vs. StrongStop. The 45 degree line indicates 
    agreement between the two estimates. This is only for the sites 
    in which both stopping rules did not reject all thresholds.}
    \label{fig:ForwardStop_vs_StrongStop}
\end{figure}

For the sites at which a threshold was selected for both the 
ForwardStop and StrongStop rules, the return level estimates 
based on the chosen threshold for each stopping rule can be 
compared. The result of this comparison can be seen in 
Figure~\ref{fig:ForwardStop_vs_StrongStop}. For a smaller 
return period (50 years), the agreement between estimates 
for the two stopping rules is quite high. This is a nice 
confirmation to have confidence in the analysis, however 
it is slightly misleading in that it does not contain the 
sites rejected by ForwardStop.

\begin{figure}[!ht]
   \centering
      \includegraphics[width=\textwidth]{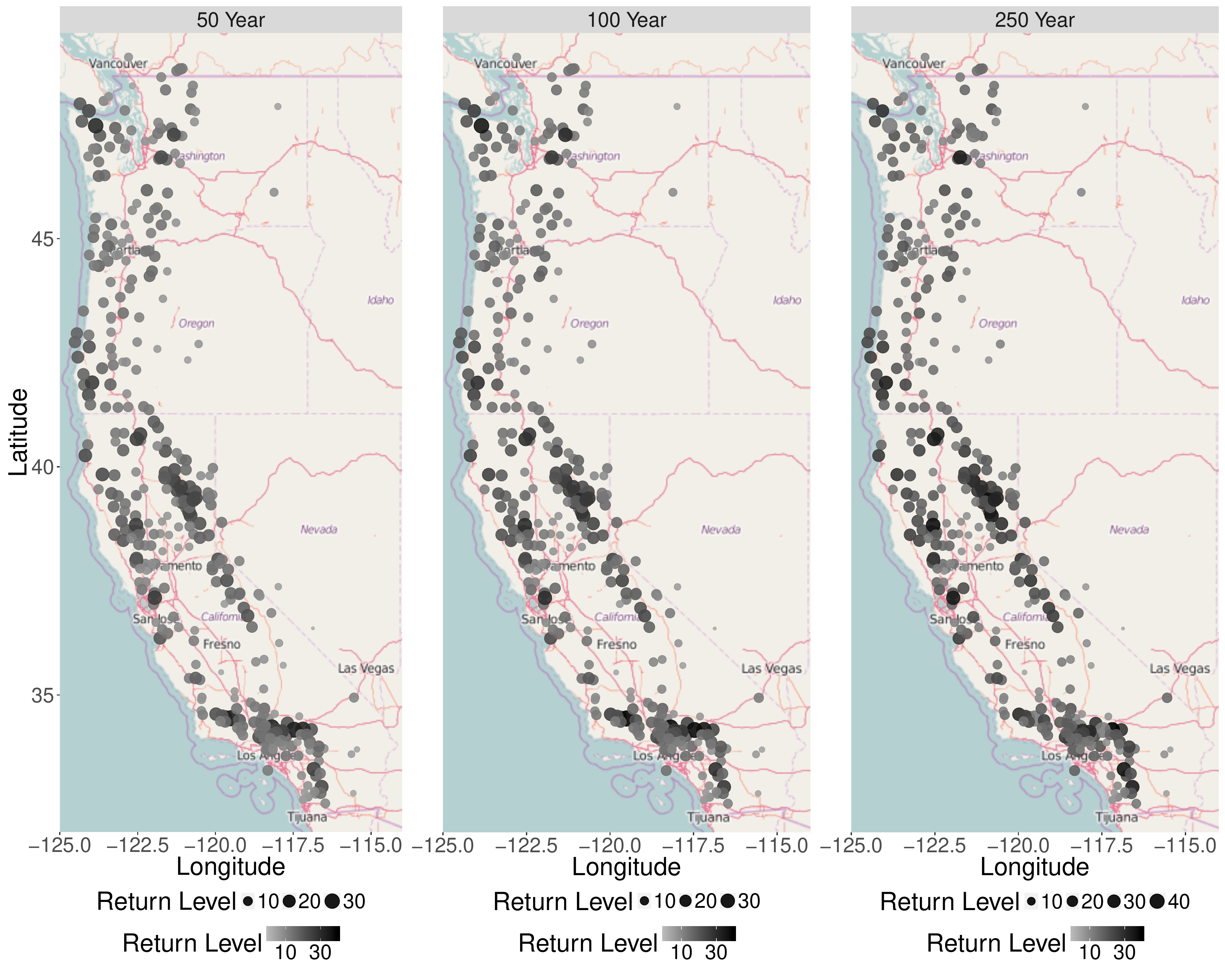}
    \caption{Map of US west coast sites with 50, 100, and 250 year 
    return level estimates via threshold selection with ForwardStop 
    and the Anderson--Darling test.}
    \label{fig:ForwardStopRL}
\end{figure}

The end result of this automated batch analysis provides a map 
of return level estimates. The 50, 100, and 250 year map 
of return level estimates from threshold selection using 
ForwardStop and the Anderson--Darling test can be seen in 
Figure~\ref{fig:ForwardStopRL}. To provide an estimate at 
any particular location in this area, some form of 
interpolation can be applied.

\section{Discussion}
\label{s:disc}

% discuss what we did, results
We propose an intuitive and comprehensive methodology for automated 
threshold selection in the peaks over threshold approach. In addition, 
it is not as computationally intensive as some competing resampling 
or bootstrap procedures~\citep{danielsson2001using, ferreira2003optimising}. 
Automation and efficiency is required when prediction using the 
peaks over threshold approach is desired at a large number of 
sites. This is achieved through sequentially testing a set of 
thresholds for goodness-of-fit to the generalized Pareto 
distribution (GPD). Previous instances of this sequential testing 
procedure did not account for the multiple testing issue. 
We apply two recently developed stopping rules~\citep{g2015sequential} 
ForwardStop and StrongStop, that control the false discovery 
rate and familywise error rate, respectively in the setting of 
(independent) ordered, sequential testing. 
It is a novel application of them to the threshold selection problem.
There is a slight caveat in our setting, that the tests are not
independent, but it is shown via simulation that these stopping rules
still provide reasonable error control here.

Four tests are compared in terms of power to detect departures from 
the GPD at a single, fixed threshold and it is found that the 
Anderson--Darling test has the most power in various non-null 
settings. \citet{choulakian2001goodness} derived the asymptotic 
null distribution of the Anderson--Darling test statistic. 
However this requires solving an integral equation. Our 
contribution, with some advice from the author, provides an 
approximate, but accurate and computationally efficient version 
of this test. To investigate the performance of the stopping rules 
in conjunction with the Anderson--Darling test, a large scale 
simulation study was conducted. Data is generated from a plausible 
distribution -- misspecified below a certain threshold and generated 
from the null GPD above. In each replicate, the bias, coverage, and 
squared error is recorded for the stopping threshold of each stopping 
rule for various parameters. The results of this simulation suggest 
that the ForwardStop procedure has on average the best performance 
using the aforementioned metrics. The entire methodology is applied 
to daily precipitation data at hundreds of sites in three U.S. west 
coast states, with the goal of creating a return level map.

% future work
Temporal or covariate varying thresholds, as discussed in 
\citet{roth2012regional} and \citet{northrop2011threshold}, is an obvious 
extension to this work. However, one particular complication 
that arises is performing model selection (i.e. what covariates 
to include), while concurrently testing for goodness-of-fit 
to various thresholds. It is clear that threshold selection 
will be dependent on the choice of model. Another possible extension 
involves testing for overall goodness-of-fit across sites 
(one test statistic). In this way, a fixed or quantile regression 
based threshold may be predetermined and then tested simultaneously 
across sites. In this setup both spatial and temporal dependence need to 
be taken into account. Handling this requires some care due to 
censoring~\citep[e.g.,][Section 2.5.2]{yan2016extremes}. In other words, 
it is not straightforward to capture the temporal dependence as 
exceedances across sites are not guaranteed to occur at the same 
points in time.

\bibliographystyle{asa}
\bibliography{GPDTests}

% AOS,AOAS: If there are supplements please fill:
%\begin{supplement}[id=suppA]
%  \sname{Supplement A}
%  \stitle{Title}
%  \slink[doi]{10.1214/00-AOASXXXXSUPP}
%  \sdatatype{.pdf}" 
%  \sdescription{Some text}
%\end{supplement}

\end{document}